\documentclass{PoS}

\leftline{\parbox{20cm}{\large BI-TP 2010/33, HU-EP-10/58, DESY 10-174, Edinburgh 2010/29, LU-ITP 2010/006, LTH 888}} 
\title{Very high order lattice perturbation theory for Wilson loops}

\ShortTitle{Very high order lattice perturbation theory for Wilson loops}

\author{R.~Horsley$^a$, G.~Hotzel$^b$,  E.-M.~Ilgenfritz$^{c,d}$,   Y.~Nakamura$^e$,
\speaker{H.~Perlt}$ ^b$,
   P.~E.~L.~Rakow$^f$, G.~Schierholz$^g$,
   A.~Schiller$^b$\\
\llap{$^a$}School of Physics, University of Edinburgh,
Edinburgh EH9 3JZ, UK\\
\llap{$^b$}Institut f\"ur Theoretische Physik, Universit\"at Leipzig,
04109 Leipzig, Germany\\
\llap{$^c$}Fakult\"at f\"ur Physik, Universit\"at Bielefeld, 
33501 Bielefeld, Germany\\
\llap{$^d$}Institut f\"ur Physik, Humboldt-Universit\"at zu Berlin, 12489 Berlin, Germany\\
\llap{$^e$}Institut f\"ur Theoretische Physik, Universit\"at Regensburg,
93040 Regensburg, Germany\\
\llap{$^f$}Theoretical Physics Division, Department of Mathematical Sciences,
University of Liverpool, Liverpool L69 3BX, UK\\
\llap{$^g$}Deutsches Elektronen-Synchrotron DESY,
22603 Hamburg, Germany\\
\\
{\rm E-mail}: perlt@itp.uni-leipzig.de}

\abstract{We calculate perturbative Wilson loops of various sizes
up to loop order $n=20$ at
different lattice sizes for pure plaquette and tree-level
improved Symanzik gauge theories
using the technique of Numerical Stochastic
Perturbation Theory.
This allows us to investigate the  behavior of
the perturbative series at high orders. 
We observe differences in the behavior of perturbative coefficients 
as a function of the loop order. 
Up to  $n=20$ we do not see evidence for the often assumed 
factorial growth of the coefficients.
Based on the observed behavior we sum this series
in a model with hypergeometric functions.
Alternatively we estimate the series in
boosted perturbation theory.
Subtracting the estimated perturbative series for the average
plaquette from the non-perturbative Monte Carlo result we estimate the gluon condensate.
}

\FullConference{The XXVIII International Symposium on Lattice Field Theory\\
		 June 14-19,2010\\
		 Villasimius, Sardinia Italy}

\begin{document}

\section{Introduction}

Since the introduction of the non-perturbative
gluon condensate by Shifman, Vainshtein and Zakharov~\cite{Shifman:1978bx} there have been many attempts to obtain 
reliable numerical results for this quantity. 
Soon it became clear that lattice gauge theory provides a promising tool to
calculate it from Wilson loops $W_{NM}$ of sizes $N \times M$. In ~\cite{Banks:1981zf,Di Giacomo:1981wt}
the plaquette was used whereas larger Wilson loops have been investigated 
in~\cite{Kripfganz:1981ri,Ilgenfritz:1982yx}. In all cases it turned out to be crucial to 
know the perturbative tail of the Wilson loops as precisely
as possible. In the last decade the application of Numerical Stochastic Perturbation Theory (NSPT)
\cite{Alfieri:2000ce} pushed the perturbative order of the plaquette up to order $n=10$
\cite{DiRenzo:2000ua} and even $n=16$~\cite{Rakow:2005yn}.

Apart from the extraction of the gluon condensate there is a general
interest in the behavior of perturbative series in QCD (for a recent investigation see~\cite{Meurice:2006cr}).
It is generally believed that these series are asymptotic, and
assumed that for large $n$ the leading growth of the coefficients $a_n$
is factorial. Using the technique of NSPT one reaches 
orders of the perturbative series where a possible set-in of this assumed
behavior can be tested - at least for finite lattices. In a more recent paper~\cite{Narison:2009ag}
Narison and Zakharov discussed the difference between short and long perturbative series and
its impact on the determination of the gluon condensate.

In order to test to what extent
$\mathcal{O}(a^2)$ 
improvement influences the behavior of the 
series we used in addition to 
the standard Wilson plaquette gauge action the  tree-level improved Symanzik
gauge action obtained by Weisz implementing Symanzik's improvement programme~\cite{Weisz:1982zw}.

We present perturbative calculations in NSPT up to order $n=20$ for
Wilson loops using the Wilson action for lattice sizes $L^4$ with $L=4,6,8,12$. 
In case of the Symanzik action we have computed the $W_{NM}$ for $L=4,6,8,10$.
The computation for $L=12$ were performed on a NEC SX-9 computer of RCNP at Osaka University,
for $L=10$ at the HLRN Berlin/Hannover,
all others on Linux/HP - clusters at  Leipzig University.

\section{NSPT calculation up to $n=20$}
\label{sec:NSPT}

NSPT allows perturbative calculations on a finite lattice up to loop order $n$
which practically cannot be reached by the standard diagrammatic approach.
A limit is set only by storage limitations and machine precision.
The algorithm is introduced and discussed in detail in~\cite{Alfieri:2000ce, Di Renzo:2004ge} -- 
we will not present it in this paper. For a detailed discussion of our results
we refer to a forthcoming paper~\cite{WLNSPT}.

\begin{figure}[h!t!b]
  \begin{center}
     \begin{tabular}{cc}
        \includegraphics[scale=0.6,clip=true]
%%         {Figures/WNML12.eps}
         {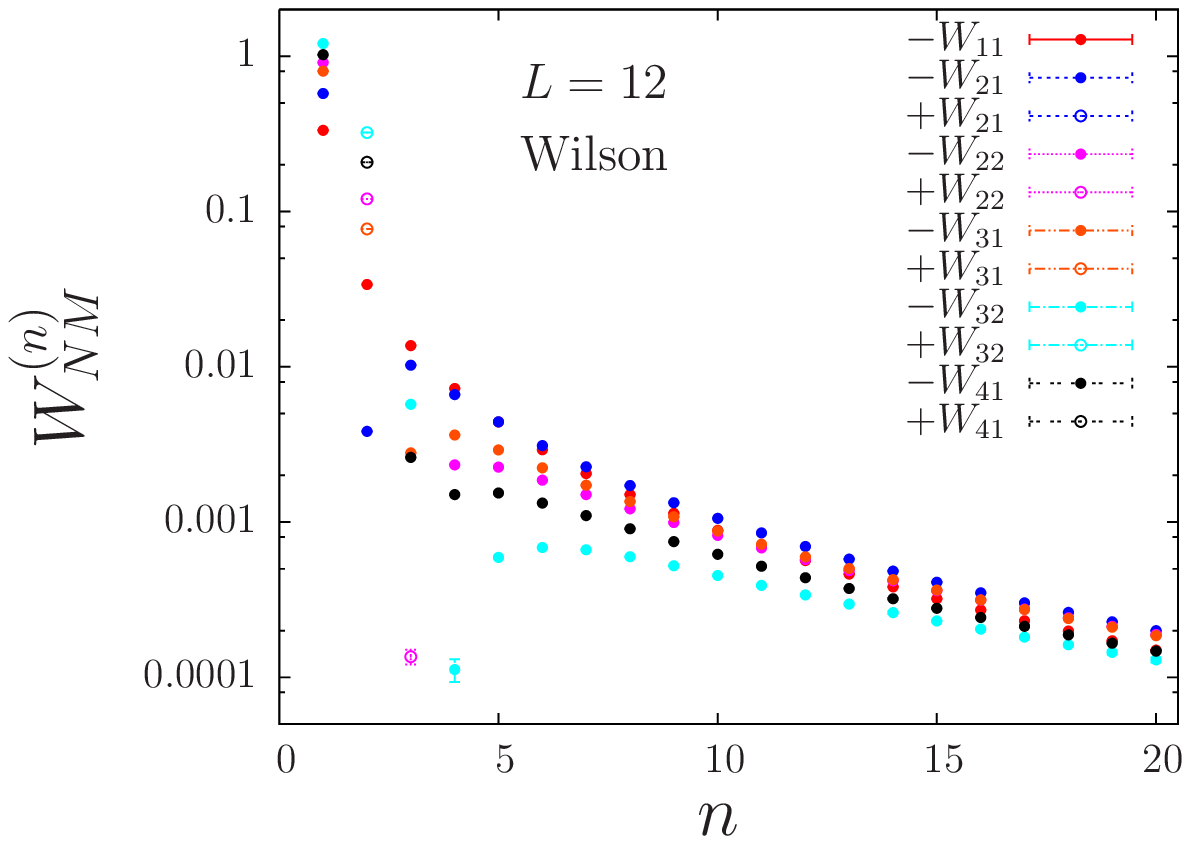}
        &
        \includegraphics[scale=0.6,clip=true]
%%         {Figures/WNML10Sym.eps}
         {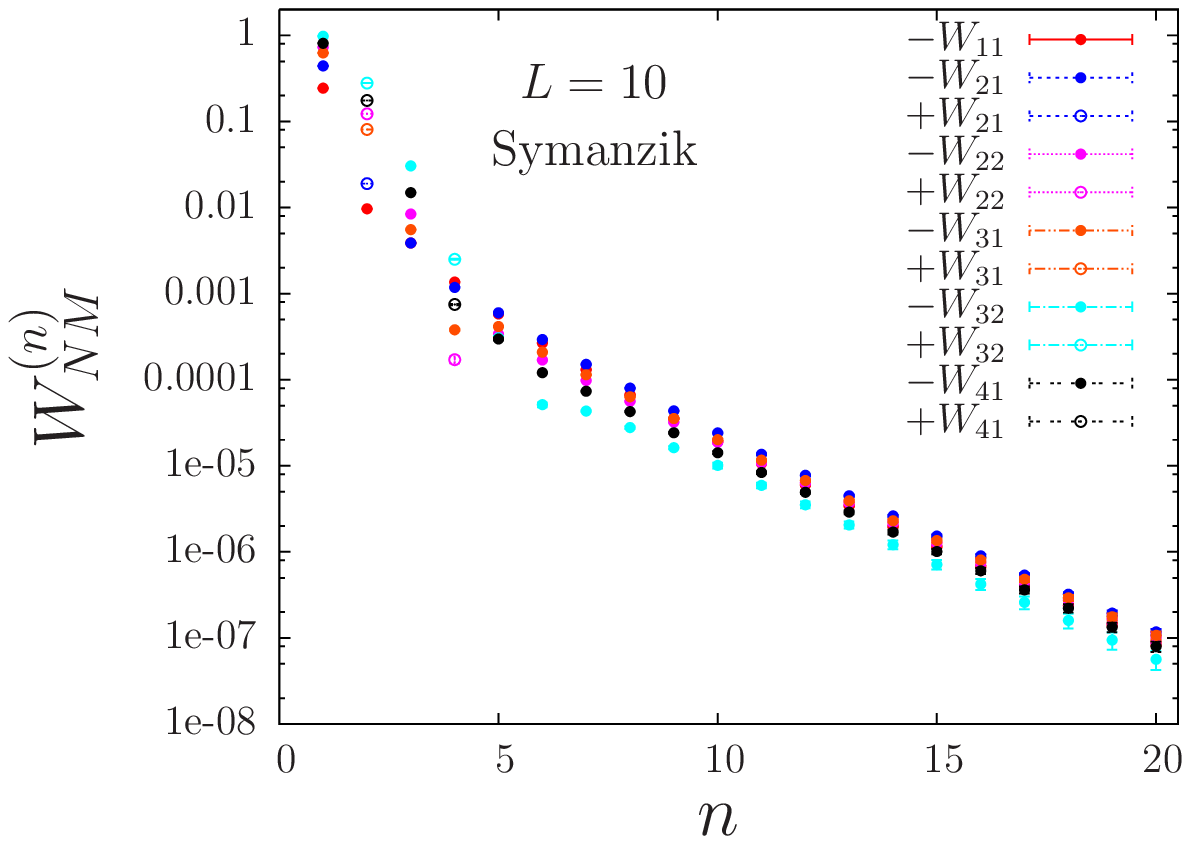}
     \end{tabular}
  \end{center}
  \caption{Selected $W_{NM}^{(n)}$ for $L=12$ and Wilson action (left) and for $L=10$ and Symanzik action (right).
 Positive/negative signs of the coefficients are given by open/full symbols.}
  \label{fig:WNMpert}
\end{figure}

In order to fix the notation we write the general expansion of a Wilson loop of size $N\times M$ in terms of the
bare lattice coupling $g$ as
\begin{equation}
W_{NM} = \sum_{n=0}^{20}\,W_{NM}^{(n)}\, g^{2n}\,.
\label{WNMPTSeries}
\end{equation}
The coefficients $W_{NM}^{(n)}$ are determined with NSPT and need to be known to
a very good precision.
In Figure \ref{fig:WNMpert} we show some results for these
coefficients using the Wilson and Symanzik actions. One can see from that Figure that the statistical errors within the NSPT algorithm are rather small.

At all $n$ all the loops behave rather similarly, with the coefficents
decreasing smoothly, in a similar way for all
loop sizes. At smaller $n$ we see that the coefficients change
sign and magnitude in a rather unpredictable way,
particularly for large loops.
We recognize some remarkable differences in the behavior between the two actions.
First, for the Symanzik action the sign change in the expansion coefficients sets in 
at smaller Wilson loop sizes $N \times M$ and is extended to larger loop orders $n$. 
Second, the values
of the Symanzik coefficients themselves are much smaller than their Wilson
action counterparts. However, concerning the relative convergence behavior of
the perturbative series (\ref{WNMPTSeries}) one should have in mind
that the coupling $g^2$ used in computing physical observables is
larger in the Symanzik case than in the Wilson case.
At larger $n$ a seemingly  asymptotic behavior (without sign changes) emerges.

\begin{figure}[h!t!b]
  \begin{center}
     \begin{tabular}{cc}
        \includegraphics[scale=0.5,clip=true]
%%         {Figures/fit_WN1_12_W.eps}
         {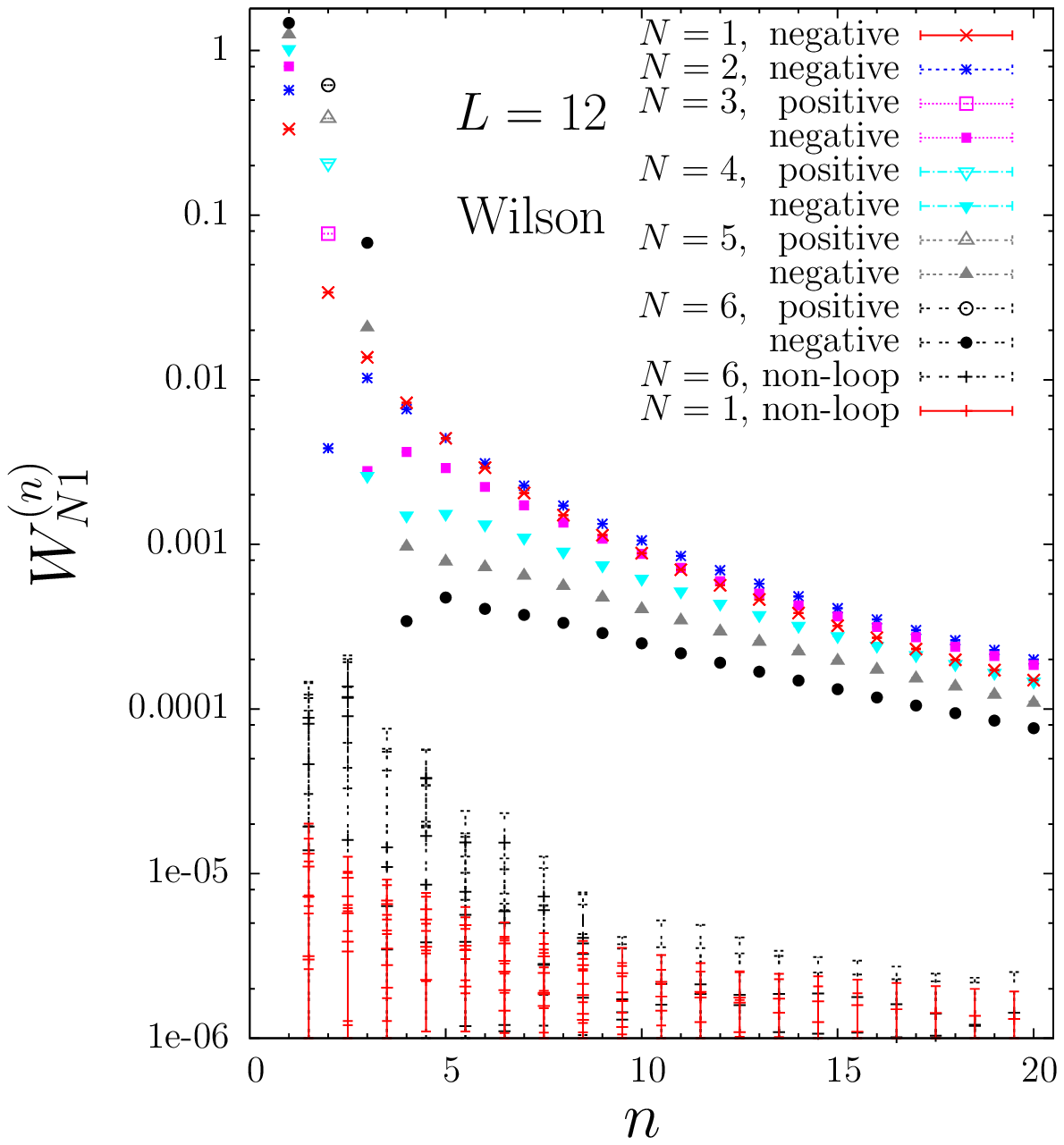}
        &
        \includegraphics[scale=0.5,clip=true]
%%         {Figures/fit_WN1_10_S.eps}
          {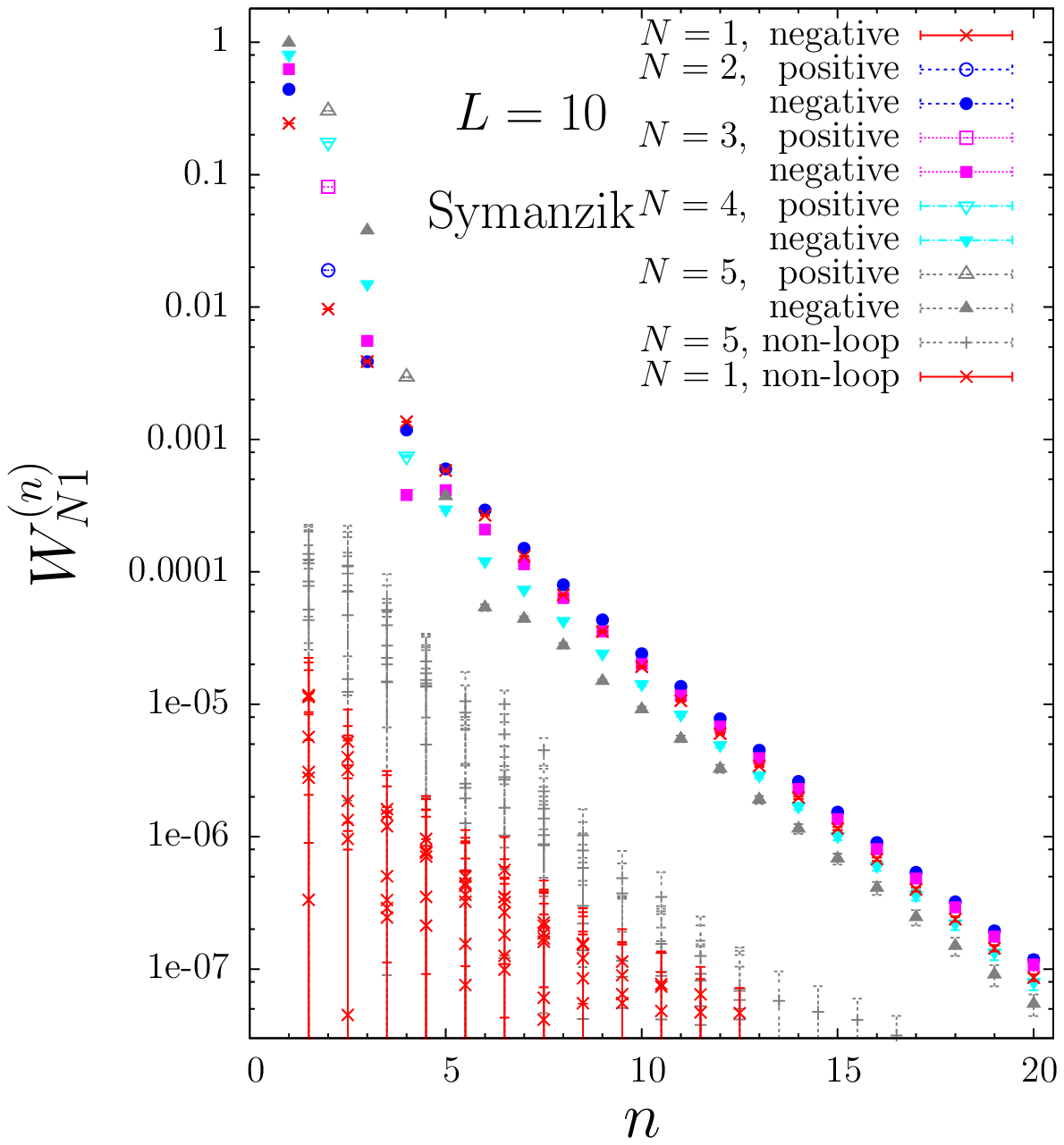}
    \end{tabular}
  \end{center}
  \caption{Selected $W_{N1}^{(n)}$ versus $n$ together with typical sizes (in magnitude) of non-loop
  contributions. Positive/negative signs of the coefficients are given by open/full symbols. Left: $L=12$ and Wilson
gauge action, right: $L=10$ and Symanzik action. Points at half-integer $n$ (the non-loop terms) are
pure noise - they show a good signal/noise ratio.}
  \label{fig:WN1SN}
\end{figure}

One essential test for the validity of our perturbative calculations 
with NSPT is the signal/noise ratio. The expansion of the (integer) loop-order $n$ is
constructed from even powers $g^{2n}$. Likewise one can
build non-loop contributions which are half-integer and combined from
odd powers of $g$ - they should vanish. In Fig. \ref{fig:WN1SN} we show that
the loop contributions are always clearly separated in magnitude from the non-loop (noise)
terms.

In addition we have to raise the question about
the infinite volume limit $L \rightarrow \infty$ of the loop series.
This is related to the definition 
of the gluon condensate which is clearly
a $L = \infty$ quantity. 
We follow the approach to extrapolate each
coefficient at finite $L$ from $W_{NM,L}^{(n)}$ to $W_{NM,\infty}^{(n)}$
using the ansatz
\begin{equation}
W_{NM,L}^{(n)} = a_{NM,n} + b_{NM,n}\, L^{-4}+ c_{NM,n}\,\log(L)\, L^{-6}\,.
\label{fitansatz}
\end{equation}
%This is a more or less heuristic point of view - neglecting possible
%intrinsic problems in course of this extrapolation, e.g.renormalons.

\section{Perturbative series at large order}
\label{sec:ptseries}

It is generally assumed that perturbative series in continuum QCD are asymptotic. 
The situation might be different for perturbative series on finite lattices.
Here we have both ultraviolet and infrared cut-offs and 
the series could be summed up to a finite value. With our computed coefficients up to order
$n=20$ we are able to check this conjecture to a so far unrivaled level.

\begin{figure}[h!t!b]
  \begin{center}
     \begin{tabular}{cc}
        \includegraphics[scale=0.72,clip=true]
%%         {Figures/CoffRatW11L12Plaq.eps}
         {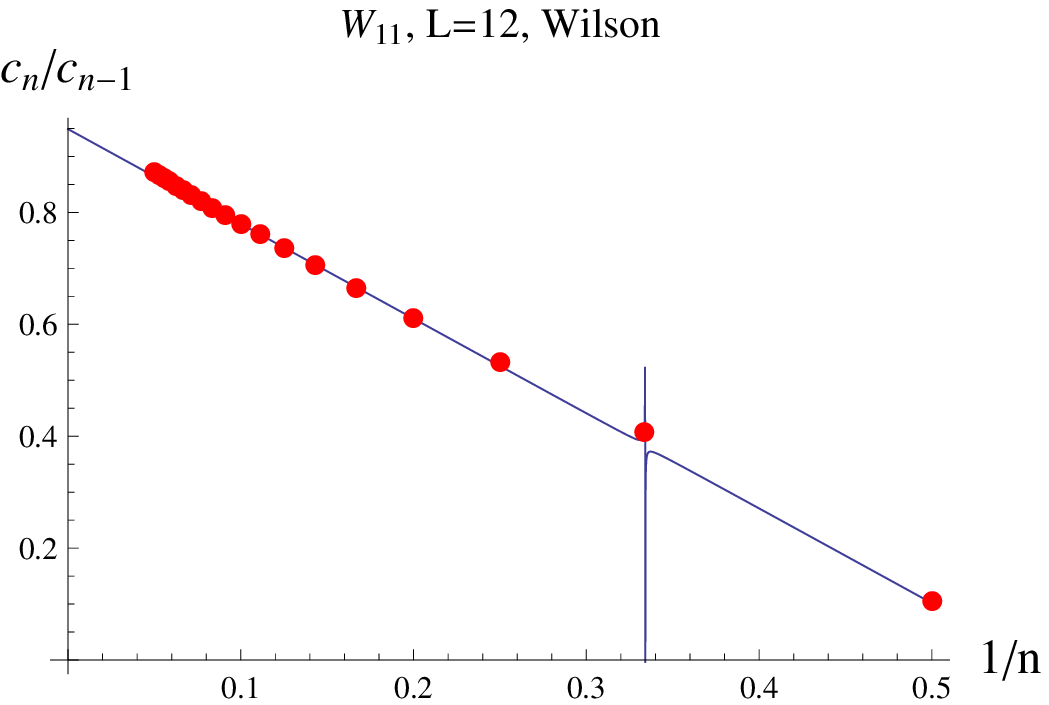}
        &
        \includegraphics[scale=0.72,clip=true]
%%         {Figures/CoffRatW11L8Symanzik.eps}
         {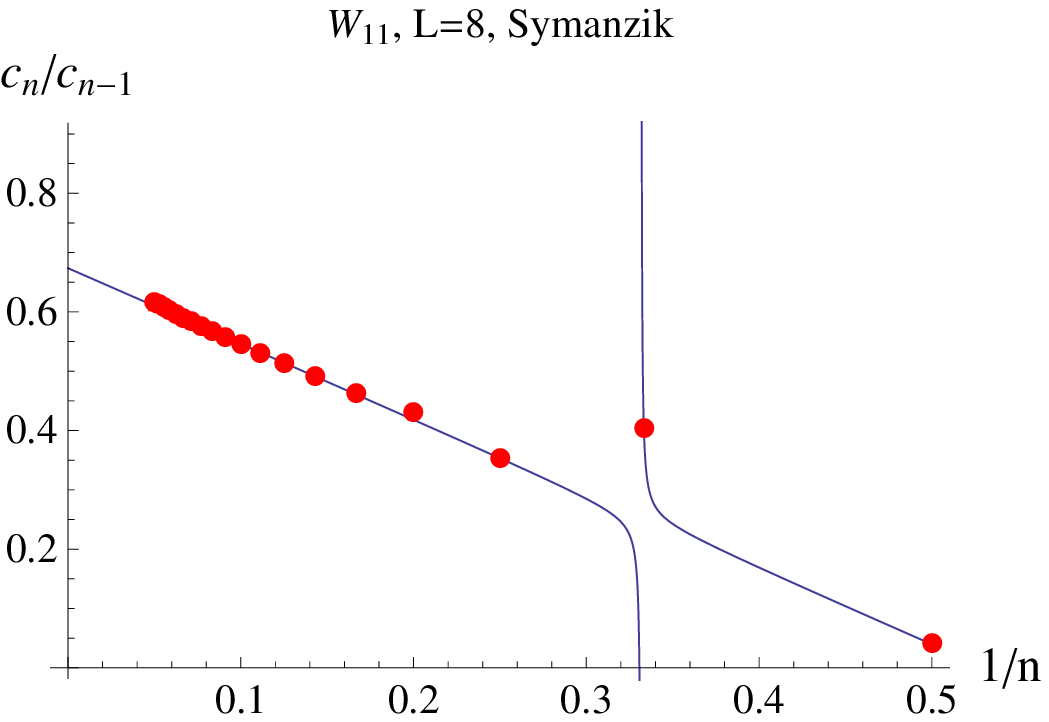}
     \end{tabular}
  \end{center}
 \caption{Domb-Sykes plots for $W_{11}$. Left: $L=12$  and Wilson action, right:  $L=8$ and Symanzik action.}
  \label{fig:DSPlots}
\end{figure}

To perform such a summation we propose a model which is a generalization of the 
ansatz used in~\cite{Horsley:2001uy}. 
Denoting the generic coefficients of the perturbative series as $c_n$
we have found that 
a large set of our data above some loop order $n_0$ can be described rather
well by the following ansatz for the ratio of subsequent coefficients
\begin{equation}
r_n=c_n/c_{n-1} = u \left(1-\frac{1+\gamma}{n}  \right) + \frac{p}{n(n+s)}\,, \quad n>n_0\,,
\label{HRSratio}
\end{equation}
where $u, \gamma, p,s$ are free parameters to be fitted.
The series based on (\ref{HRSratio}) can be summed up to $n=\infty$
for $g^2 < 1/u$. For $n_0=1$ we obtain as solution
\begin{eqnarray}
W_{NM,\infty}&=& \frac{(s+1) \,W_{NM}^{(1)}\,\, _2F_1\left((1+s-\gamma)/2-\tau,(1+s-\gamma)/2+\tau;s+1;g^2\, u\right)}{p-(s+1)\, u \,\gamma }\,,
\label{finsum1}\\
\tau &=& \frac{1}{2}\,\sqrt{\left((\gamma+s+1)^2 u-4 p\right)/u}\nonumber\,.
\end{eqnarray}
As final hypergeometric model at $n_0=1$ we use
\begin{equation}
W_{NM,HYP}= W_{NM,\infty} - 
(c_{1,HYP}-W_{NM}^{(1)})\,g^2\,,
\label{finsum2}
\end{equation}
where $c_{1,HYP}$ is the coefficient of $g^2$ in the corresponding $g^2$-expansion of $W_{NM,\infty}$.
For other values of $n_0$ the formulae (\ref{finsum1}) and (\ref{finsum2}) change accordingly.
In Fig. \ref{fig:DSPlots} we show two example plots for the ratio $r_n$ together with
the  model fit based on (\ref{HRSratio}). In all investigated cases we do not observe
a factorial growth of the coefficients up to order $n=20$.
Formally, one can use  model (\ref{finsum2}) (used for finite lattices) also for the
$L \rightarrow \infty$ extrapolated coefficients.
%%Although the extrapolation seems to yield smooth limits, it is certainly not allowed to sum up the series 
%%based on these coefficients up to infinity. 

\begin{figure}[h!t!b]
  \begin{center}
     \begin{tabular}{cc}
        \includegraphics[scale=0.62,clip=true]
%%         {Figures/W11SumPlot.eps}
         {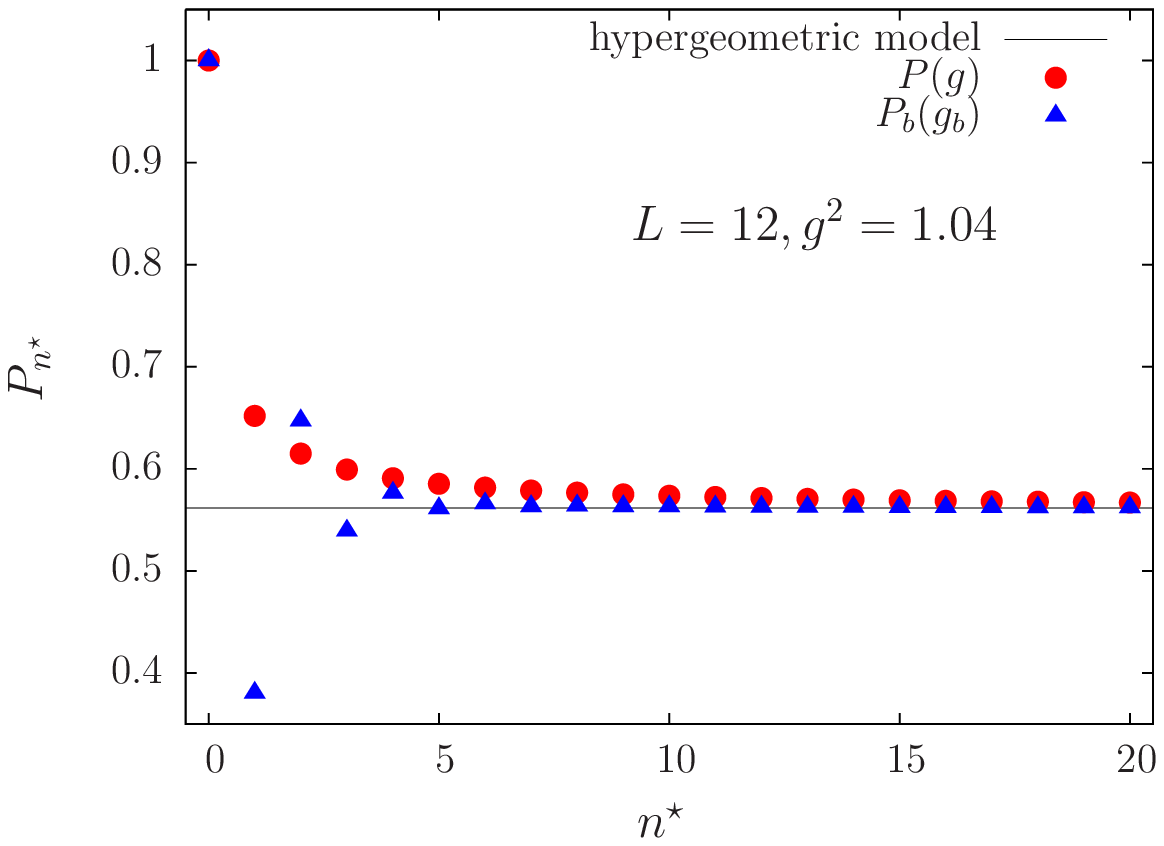}
        &
        \includegraphics[scale=0.62,clip=true]
%%         {Figures/W11SumPlot8_21.eps}
          {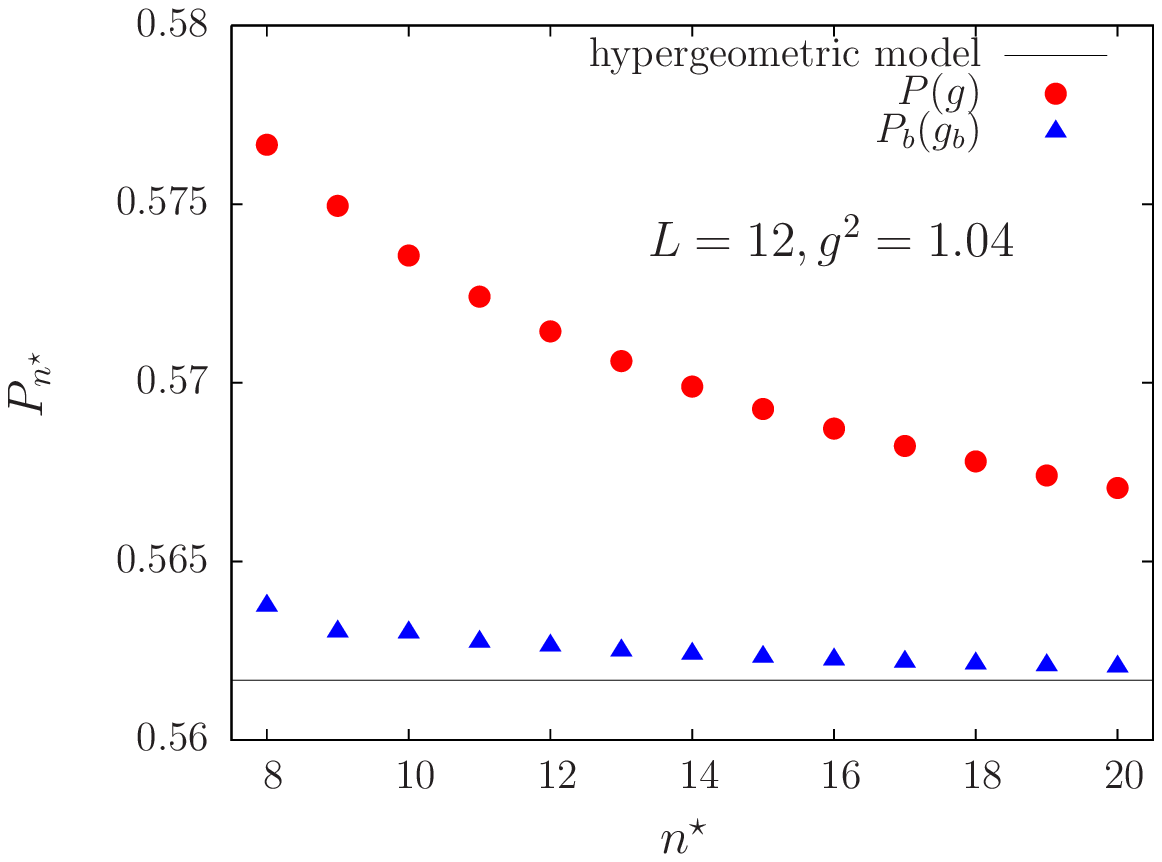}
    \end{tabular}
  \end{center}
  \caption{Perturbative plaquette $P_{n^\star}$ as function of loop order $n^\star$ for naive and boosted series; 
           left: all orders, right: zoomed in by the choice $8 \le n^\star \le 20$. The choice $g^2=1.04$ corresponds
%%to the largest coupling where the solution (\ref{finsum1}) remains real.}
to the largest coupling where the solution (3.2) remains real.}
  \label{fig:W11SumCompare}
\end{figure}

A possible alternative method consists in applying boosted perturbation theory, i.e. a rearrangement of
the series in terms of a boosted coupling
$g_b$~\cite{Rakow:2005yn} ($W_{11} \equiv P$).
Using the summed perturbative series 
$P_{n^\star}(g^2)= 1+\sum_{n=1}^{n^\star}\,W_{11}^{(n)}\,g^{2n}$ 
%As a result we get an almost  plateau of the sum as function
%of perturbative order which is used as the final result~\cite{Rakow:2005yn}
we get the boosted summed plaquette $P_{20,b}(g_{b})$ as follows
\begin{equation}
g^2 \rightarrow g^2_{b}=\frac{g^2}{P_{20}(g)}\quad
\rightarrow \quad P_{20,b}(g_{b})=1+\sum_{n=1}^{20}\,W_{b,11}^{(n)}\,g^{2n}_{b}\,.
\label{boost}
\end{equation}
In Fig. \ref{fig:W11SumCompare} we compare the summed perturbative series 
$P_{n^\star}(g)$ with $P_{n^\star,b}(g_{b})$ as function of maximal
loop order $n^\star$ for $L=12$.
One clearly recognizes a distinct plateau
reached by boosting for $n^\star<20$. The naive series is far from the model value computed from (\ref{finsum2}). 
Although this behavior of the boosted perturbation theory is found for finite $L$,
we assume that it remains valid also in the infinite volume limit.

\vspace{-1cm}
\section{Non-perturbative gluon condensate}

Having under control the large order perturbative
series we estimate now the non-perturbative gluon
condensate $\langle (\alpha_s/\pi) GG \rangle$.
Being a physical quantity we should use 
the $L=\infty$ extrapolated values of the perturbative coefficients.
On the other hand it could be useful to get numbers for
the "gluon condensate" at finite $L$ also: they provide some
numerical insight how the infinite volume limit is approached.

We use the relation between the 
plaquette measured in Monte Carlo simulations $P_{MC}$ and its perturbative analogue $P$:
\begin{equation}
  P_{MC} = P + a^4 \frac{\pi^2}{36}\left(\frac{b_0 g^2}{\beta(g)}  \right) \langle 
  \frac{\alpha_s}{\pi} GG \rangle \,,
  \label{GG1}
\end{equation}
which defines the gluon condensate.
In (\ref{GG1}) $\beta(g)$ denotes the standard $\beta$-function with $b_0$ being its
leading coefficient.
{}From (\ref{GG1}) it is obvious that the precision of the difference $\Delta P=P- P_{MC}$
determines the reliability of an estimate for $\langle \frac{\alpha_s}{\pi} GG \rangle$.

\begin{figure}[h!t!b]
  \begin{center}
     \begin{tabular}{cc}
        \includegraphics[scale=0.61,clip=true]
%%         {Figures/C4C2Boosted.eps}
          {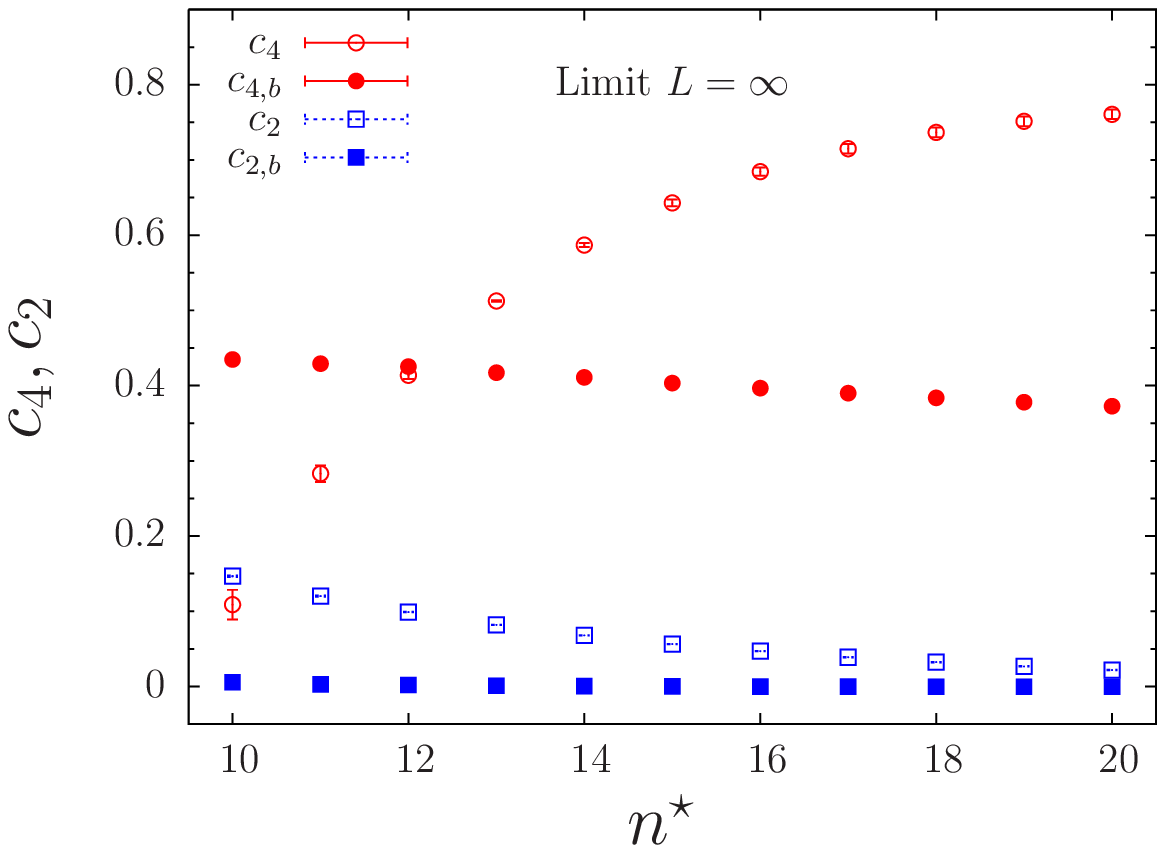}
       &
        \includegraphics[scale=0.61,clip=true]
%%         {Figures/diffPaaBoosted.eps}
         {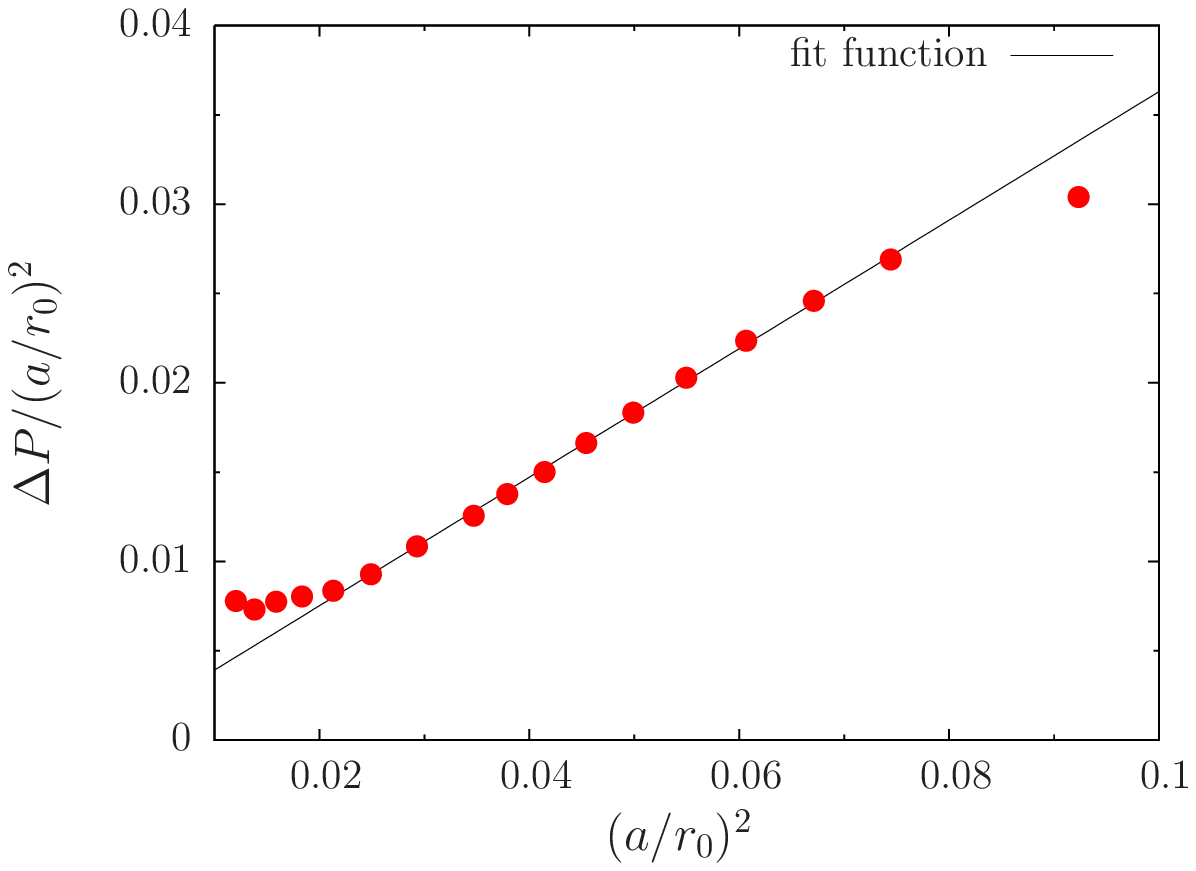}
     \end{tabular}
  \end{center}
  \caption{Left: Comparison of $c_4(n^\star)$ and $c_2(n^\star)$  for naive and boosted ($b$)
           perturbation theory in the extrapolated limit $L = \infty$.
Right: $\Delta P/(a/r_0)^2$ as function of $(a/r_0)^2$ for boosted
%%           perturbation theory using $n^\star = 20$ in the infinite volume limit. The  line
           perturbation theory. The  line
           is a linear fit in $a^2$  for the range $0.02 \le (a/r_0)^2 \le 0.08$,
$r_0$ denotes the Sommer scale.}
  \label{fig:W11Sum}
\end{figure}

It has been proposed~\cite{Narison:2009ag} to extract 
the gluon condensate taking into account a perturbative series of the
plaquette summed to an order $n^\star$ ($P_{n^\star}$) from the following ansatz
\begin{equation}
  a^4 \frac{\pi^2}{36}\left(\frac{-b_0 g^2}{\beta(g)}  \right) \langle \frac{\alpha_s}{\pi} GG \rangle =\Delta P_{n^\star} = 
P_{n^\star} - P_{MC}\,,
  \label{GG11}
\end{equation}
where $\Delta P_{n^\star}$ contains possible $a^2$ and $a^4$ contributions
\begin{equation}
  \Delta  P_{n^\star} = c_2(n^\star) \,a^2+ c_4(n^\star)\, a^4\,.
\label{GG12}
\end{equation}
The authors in~\cite{Narison:2009ag} argued that non-zero values of the coefficient $c_2(n^\star)$ are an
artefact due to a truncation of the series: above some value of
$n^\star$ this coefficient should vanish. Our perturbative series for both, the naive series (finite $L$) and
boosted series ($L=\infty$) confirm their hypothesis: for boosted perturbation theory the
$a^2$-term decreases rapidly to zero for $n^\star > 10$ (see left of Fig. \ref{fig:W11Sum}); for naive
perturbation theory and finite $L$ one has to sum the series to much larger $n^\star$.

{}From the results obtained so far we conclude 
 that the most reliable
estimate for the gluon condensate can be obtained for the Wilson
action using boosted perturbation theory.
In the right of Figure \ref{fig:W11Sum} we show ${\Delta P}/{(a/r_0)^2}$ as function of $(a/r_0)^2$
using boosted perturbation theory with $n^\star=20$ and
our own non-perturbative plaquette measurements at different $\beta$ values and
lattice sizes for $r_0/a \ge 2.5$.
We clearly recognize a dominant linear $a^2$ behavior.
The necessary conversion $\beta \leftrightarrow (a/r_0)$ is accomplished for
the plaquette action by the Necco-Sommer relation~\cite{Necco:2001xg}.

In Table \ref{GGtab} we give numbers for the quantity $\langle ({\alpha_s}/{\pi})G\,G \rangle$ for {\sl finite}
$L$ and boosted perturbation theory at $L=\infty$. For the perturbative values at finite $L$
we use the hypergeometric model as given in (\ref{finsum2}). 
The table shows that the boosted result for extrapolation $L=\infty$ 
agrees within errors with the model value at finite lattice size  $L=12$. 
The given errors have been estimated by varying the fit window $\beta_{\min} \le \beta \le \beta_{\max}$.
\begin{table}[!htb]
\vspace{0.5cm}
\begin{center}
\begin{tabular}{|c  |c |c|c|}
\hline
$L$ &  $r_0^4\,\langle \frac{\alpha_s}{\pi}G\,G \rangle$ &  $\langle \frac{\alpha_s}{\pi}G\,G \rangle$ [GeV$^4$]&
fit range\\
 \hline
6                  & $1.08(10)$   & $0.034(3)$ & $5.73 \le \beta \le 6.00$\\
8                  & $1.22(11)$   & $0.039(4)$ & $5.78 \le \beta \le 6.10$ \\
12                 & $1.29(17)$   & $0.041(6)$ & $5.78 \le \beta \le 6.27$ \\
\hline
$\infty$, boosted  & $1.33(7)$   & $0.042(2)$ & $5.78 \le \beta \le 6.17$ \\
\hline
\end{tabular}
\end{center}
\caption{Gluon condensate at finite $L$ for Wilson gauge action ($r_0=0.467$ fm). The
last column shows the used $\beta$-fit range.}
\label{GGtab}
\end{table}
Our reference value for the gluon condensate is that for $L=\infty$ and boosted
perturbation theory given in Table \ref{GGtab}. With the chosen $r_0$ it is
larger than the value obtained by sum rule calculations~\cite{Shifman:1978bx}.

\section*{Acknowledgements}
This investigation has been supported partly by DFG under contract SCHI 422/8-1.
We  thank the RCNP at Osaka university and the HLRN Berlin/Hannover
for providing computer resources.


\begin{thebibliography}{99}
%\cite{Shifman:1978bx}
\bibitem{Shifman:1978bx}
M.~A.~Shifman, A.~I.~Vainshtein and V.~I.~Zakharov,
  %``QCD And Resonance Physics. Sum Rules,''
  \emph{Nucl.\ Phys.\  B} {\bf 147} (1979) 385.
  %%CITATION = NUPHA,B147,385;%%

%\cite{Banks:1981zf}
\bibitem{Banks:1981zf}
  T.~Banks, R.~Horsley, H.~R.~Rubinstein and U.~Wolff,
  %``Estimate Of The Gluon Condensate From Monte Carlo Calcul
  \emph{Nucl.\ Phys.\  B} {\bf 190} (1981) 692.


%\cite{Di Giacomo:1981wt}
\bibitem{Di Giacomo:1981wt}
  A.~Di Giacomo and G.~C.~Rossi,
  %``Extracting The Vacuum Expectation Value Of The Quantity Alpha / Pi G G From
  %Gauge Theories On A Lattice,''
  \emph{Phys.\ Lett\  B} {\bf 100} (1981) 481.
  %%CITATION = PHLTA,B100,481;%%

%\cite{Kripfganz:1981ri}
\bibitem{Kripfganz:1981ri}
  J.~Kripfganz,
  %``Gluon Condensate From SU(2) Lattice Gauge Theory,''
  \emph{Phys.\ Lett.\  B} {\bf 101} (1981) 169; R.~Kirschner, J.~Kripfganz, J.~Ranft and A.~Schiller,
  %``Short Distance Expansion Of Wilson Loops, Gluon Condensation And Monte
  %Carlo Lattice Results,''
  \emph{Nucl.\ Phys.\  B} {\bf 210} (1982) 567.


%\cite{Kirschner:1982qg}
%\bibitem{Kirschner:1982qg}
%  R.~Kirschner, J.~Kripfganz, J.~Ranft and A.~Schiller,
  %``Short Distance Expansion Of Wilson Loops, Gluon Condensation And Monte
  %Carlo Lattice Results,''
%  Nucl.\ Phys.\  B {\bf 210} (1982) 567.
  %%CITATION = NUPHA,B210,567;%%


%\cite{Ilgenfritz:1982yx}
\bibitem{Ilgenfritz:1982yx}
  E.-M.~Ilgenfritz and M.~M\"uller-Preussker,
  %``SU(3) Gluon Condensate From Lattice Mc Data,''
  Phys.\ Lett.\  B {\bf 119} (1982) 395.
  %%CITATION = PHLTA,B119,395;%%


%\cite{Alfieri:2000ce}
\bibitem{Alfieri:2000ce}
  R.~Alfieri, F.~Di Renzo, E.~Onofri and L.~Scorzato,
  %``Understanding stochastic perturbation theory: Toy models and  statistical
  %analysis,''
  \emph{Nucl.\ Phys.\  B} {\bf 578} (2000) 383
[\href{http://arxiv.org/abs/hep-lat/0002018}{\tt arXiv:hep-lat/0002018}].
%%  [arXiv:hep-lat/0002018].
  %%CITATION = NUPHA,B578,383;%%

%\cite{DiRenzo:2000ua}
\bibitem{DiRenzo:2000ua}
  F.~Di Renzo and L.~Scorzato,
  %``A consistency check for renormalons in lattice gauge theory:  beta**(-10)
  %contributions to the SU(3) plaquette,''
  \emph{JHEP} {\bf 0110} (2001) 038
[\href{http://arxiv.org/abs/hep-lat/0011067}{\tt arXiv:hep-lat/0011067}].
%%  [arXiv:hep-lat/0011067].
  %%CITATION = JHEPA,0110,038;%%

%\cite{Rakow:2005yn}
\bibitem{Rakow:2005yn}
  P.~E.~L.~Rakow,
  %``Stochastic perturbation theory and the gluon condensate,''
  PoS {\bf LAT2005} (2006) 284
[\href{http://arxiv.org/abs/hep-lat/0510046}{\tt arXiv:hep-lat/0510046}].
%%  [arXiv:hep-lat/0510046].
  %%CITATION = POSCI,LAT2005,284;%%

%\cite{Meurice:2006cr}
\bibitem{Meurice:2006cr}
  Y.~Meurice,
  %``The non-perturbative part of the plaquette in quenched QCD,''
  \emph{Phys.\ Rev.\  D} {\bf 74} (2006) 096005
[\href{http://arxiv.org/abs/hep-lat/0609005}{\tt arXiv:hep-lat/0609005}].
%%  [arXiv:hep-lat/0609005].
  %%CITATION = PHRVA,D74,096005;%%

%\cite{Burgio:1997hc}
%\bibitem{Burgio:1997hc}
%  G.~Burgio, F.~Di Renzo, G.~Marchesini and E.~Onofri,
  %``Lambda**2-contribution to the condensate in lattice gauge theory,''
%  \emph{Phys.\ Lett.\  B} {\bf 422} (1998) 219
%[\href{http://arxiv.org/abs/hep-ph/9706209}{\tt arXiv:hep-ph/9706209}].
%%  [arXiv:hep-ph/9706209].
  %%CITATION = PHLTA,B422,219;%%


%\cite{ZinnJustin}
%\bibitem{ZinnJustin}
%J.~C.~LeGuillou and J.~Zinn-Justin, \emph{Large-Order Behavior
%of Perturbation Theory}, North-Holland, Amsterdam,
%1990.


%\cite{Narison:2009ag}
\bibitem{Narison:2009ag}
  S.~Narison and V.~I.~Zakharov,
  %``Duality between QCD Perturbative Series and Power Corrections,''
  Phys.\ Lett.\  B {\bf 679} (2009) 355
[\href{http://arxiv.org/abs/0906.4312}{\tt arXiv:0906.4312[hep-ph]}].
%  [arXiv:0906.4312 [hep-ph]].
  %%CITATION = PHLTA,B679,355;%%

%\cite{Weisz:1982zw}
\bibitem{Weisz:1982zw}
  P.~Weisz,
  %``Continuum Limit Improved Lattice Action For Pure Yang-Mills Theory. 1,''
  Nucl.\ Phys.\  B {\bf 212} (1983) 1.
  %%CITATION = NUPHA,B212,1;%%

%\cite{Di Renzo:2004ge}
\bibitem{Di Renzo:2004ge}
  F.~Di Renzo and L.~Scorzato,
  %``Numerical stochastic perturbation theory for full QCD,''
  \emph{JHEP} {\bf 0410}, 073 (2004)
[\href{http://arxiv.org/abs/hep-lat/0410010}{\tt arXiv:hep-lat/0410010}].
%%  [arXiv:hep-lat/0410010].
  %%CITATION = JHEPA,0410,073;%%

%\cite{WLNSPT}
\bibitem{WLNSPT}
  R.~Horsley et al., in preparation.
  %``One Loop Perturbative Calculation Of Wilson Loops On Finite Lattices,''
 % Nucl.\ Phys.\  B {\bf 251} (1985) 254.
  %%CITATION = NUPHA,B251,254;%%


%\cite{Horsley:2001uy}
\bibitem{Horsley:2001uy}
  R.~Horsley, P.~E.~L.~Rakow and G.~Schierholz,
  %``Separating perturbative and non-perturbative contributions to the
  %plaquette,''
  \emph{Nucl.\ Phys.\ Proc.\ Suppl.}\  {\bf 106} (2002) 870
[\href{http://arxiv.org/abs/hep-lat/0110210}{\tt arXiv:hep-lat/0110210}].
%  [arXiv:hep-lat/0110210].
  %%CITATION = NUPHZ,106,870;%%

%\cite{Dosch:1988ha}
%\bibitem{Dosch:1988ha}
%  H.~G.~Dosch and Yu.~A.~Simonov,
  %``The Area Law of the Wilson Loop and Vacuum Field Correlators,''
%  Phys.\ Lett.\  B {\bf 205} (1988) 339.
  %%CITATION = PHLTA,B205,339;%%


%\cite{Necco:2001xg}
\bibitem{Necco:2001xg}
  S.~Necco and R.~Sommer,
  %``The N(f) = 0 heavy quark potential from short to intermediate  distances,''
  \emph{Nucl.\ Phys.\  B} {\bf 622} (2002) 328
[\href{http://arxiv.org/abs/hep-lat/0108008}{\tt arXiv:hep-lat/0108008}].
%%  [arXiv:hep-lat/0108008].
  %%CITATION = NUPHA,B622,328;%%

%\cite{ISP}
%\bibitem{ISP}
%  E.-M.~Ilgenfritz, H.~Perlt, A.~Schiller,
%  in preparation.


\end{thebibliography}
\end{document}